\documentclass[iop,apj]{emulateapj}
\usepackage{CJK}
\usepackage{natbib}
\usepackage[dvips]{color}
\usepackage{multirow}
\usepackage{amsmath}
\usepackage{amsfonts}
\newcommand{\NII}{[N\,{\sc ii}]}
\newcommand{\CII}{[C\,{\sc ii}]}
\newcommand{\CI}{[C\,{\sc i}]}
\newcommand{\LNII}{$L_{\rm [N\,{\scriptsize \textsc{ii}}]}$}

\newcommand{\LIR}{$L_{\rm IR}$}
\newcommand{\LFIR}{$L_{\rm FIR}$}

\newcommand{\NIIIO}{[N\,{\sc ii}] 205 $\mu$m}

\newcommand{\fircolor}{$f_{60}/f_{100}$}

\newcommand{\mum}{$\mu$m}

\newcommand{\kms}{km\,s$^{-1}$}
\newcommand{\midco}{CO~(7$-$6)}
\newcommand{\lco}{$L_{\rm CO\,(7-6)}$}
\newcommand{\lessj}{LESS~073}
\begin{document}
\shortauthors{Zhao et al.}
\title{ALMA Imaging of the \midco\ Line Emission in the Submillimeter Galaxy \lessj\ at redshift 4.755$^\star$}
\author{Yinghe Zhao\altaffilmark{1,2,3}, Nanyao Lu\altaffilmark{4, 5}, Tanio D\'{i}az-Santos\altaffilmark{6}, Yu Gao\altaffilmark{7}, C. Kevin Xu\altaffilmark{4, 5}, Vassilis Charmandaris\altaffilmark{8, 9}, Hanae Inami\altaffilmark{10}, Dimitra Rigopoulou\altaffilmark{11}, David B. Sanders\altaffilmark{12}, Jiasheng Huang\altaffilmark{4, 5}, and Zhong Wang\altaffilmark{4, 5}}
\altaffiltext{$\star$}{The National Radio Astronomy Observatory is a facility of the National Science Foundation operated under cooperative agreement by Associated Universities, Inc.}
\altaffiltext{1}{Yunnan Observatories, Chinese Academy of Sciences, Kunming 650011, China; zhaoyinghe@ynao.ac.cn}
\altaffiltext{2}{Key Laboratory for the Structure and Evolution of Celestial Objects, Chinese Academy of Sciences, Kunming 650011, China}
\altaffiltext{3}{Center for Astronomical Mega-Science, CAS, 20A Datun Road, Chaoyang District, Beijing 100012, China}
\altaffiltext{4}{National Astronomical Observatories of China, Chinese Academy of Sciences, Beijing 100012, China}
\altaffiltext{5}{China-Chile Joint Center for Astronomy, Chinese Academy of Sciences, Camino El Observatorio, 1515 Las Condes, Santiago, Chile}
\altaffiltext{6}{N\'{u}cleo de Astronom\'{i}a de la Facultad de Ingenier\'{i}a, Universidad Diego Portales, Av. Ej\'{e}rcito Libertador 441, Santiago, Chile}
\altaffiltext{7}{Department of Astronomy, Xiamen University, Xiamen, Fujian 361005, China}
\altaffiltext{8}{Department of Physics, University of Crete, GR-71003 Heraklion, Greece}
\altaffiltext{9}{IAASARS, National Observatory of Athens, GR-15236, Penteli, Greece}
\altaffiltext{10}{Hiroshima Astrophysical Science Center, Hiroshima University, 1-3-1 Kagamiyama, Higashi-Hiroshima, Hiroshima 739-8526, Japan}
\altaffiltext{11}{Department of Physics, University of Oxford, Keble Road, Oxford OX1 3RH, UK}
\altaffiltext{12}{University of Hawaii, Institute for Astronomy, 2680 Woodlawn Drive, Honolulu, HI 96822, USA}

\date{Received:~ Accepted:~}

\begin{abstract}
In this paper we present our imaging observations on the \midco\ line and its underlying continuum emission of the young submillimeter galaxy \lessj\ at  redshift 4.755, using the Atacama Large Millimeter/submillimeter Array (ALMA). At the achieved resolution of $\sim$$1\arcsec.2\times0\arcsec.9$ ($8\times6$~kpc$^2$), the \midco\ emission is largely unresolved (with a deconvolved size of $1\arcsec.1(\pm0\arcsec.5)\times0\arcsec.9(\pm0\arcsec.8)$.), and the continuum emission is totally unresolved. The \midco\ line emission has an integrated flux of $0.86\pm0.08$~Jy~\kms, and a line width of $343\pm40$~\kms. The continuum emission has a flux density of 0.51~mJy. By fitting the observed far-infrared (FIR) spectral energy distribution of \lessj\ with a single-temperature modified blackbody function, we obtained a dust temperature $T_{\rm dust}=57.6\pm3.5$~K, 60-to-100~\mum\ flux density ratio $f_{60}/f_{100}=0.86\pm0.08$, and total infrared luminosity $L_{\rm IR}=(5.8\pm0.9)\times10^{12}~L_\odot$. The SED-fit-based \fircolor\ is consistent with those estimated from various line ratios as advocated by our earlier work, indicating that those proposed line-ratio-based method can be used to practically derive \fircolor\ for high-$z$ sources. The total molecular gas mass of \lessj\ is $(3.3\pm1.7)\times10^{10}~M_\odot$, and the inferred gas depletion time is about 43~Myr. 
\end{abstract}

\keywords{galaxies: active --- galaxies: nuclei --- galaxies: ISM --- galaxies: starburst --- galaxies: evolution --- submillimeter: galaxies}

\section{Introduction}
Star formation (SF) transforms gas into stars and thus is one of the most fundamental drivers of galaxy evolution. It is crucial to have an effective way to derive star formation rates (SFRs) for galaxies spanning a large range of look-back times in order to understand galaxy evolution leading back to initial conditions. Currently, SFRs have been inferred using continuum or spectral line emission from a wide range of wavelengths (e.g., Calzetti et al. 2009 and references therein; Kennicutt \& Evans 2012; Zhao et al. 2013, 2016; De Looze et al. 2014; Sargsyan et al. 2014; Daddi et al. 2015; Liu et al. 2015; Lu et al. 2015, hereafter Lu15). Among these SFR indicators, the ones (e.g., Zhao et a. 2013; De Looze et al. 2014; Sargsyan et al. 2014; Lu15) recently calibrated with far-infrared (FIR) emission lines are of particular interest since not only they are less affected by dust extinction but they can also be followed up by the Atacama Large Millimeter/submillimeter Array (ALMA; Wootten \& Thompson 2009) over a large redshift range.

Based on a sample of local luminous infrared galaxies (LIRGs; $L_{\rm IR} \equiv L(8-1000\,\mu{\rm m)} >10^{11} L_\odot$) observed with the {\it Herschel Space Observatory} (Pilbratt et al. 2010), Lu15 proposed that the combination of \midco\ (rest-frame 806.652 GHz) and either the \CII\ 158 \mum\ (1900.56 GHz; hereafter \CII) or \NIIIO\ line (1461.134 GHz; hereafter \NII) can be used to simultaneously estimate the SFR and FIR color (i.e., the rest-frame 60-to-100 \mum\ flux ratio, hereafter \fircolor) in high-$z$ galaxies. For local LIRGs and ultra-LIRGs (ULIRGs; $L_{\rm IR} >10^{12} L_\odot$), SFRs inferred from the \midco\ line luminosity (\lco) have an accuracy of 30\%, irrespective of whether the galaxy hosts an active galactic nucleus (AGN) or not (Lu et al. 2014, 2015, 2017a), whereas SFRs derived with the \CII\ (for normal star-forming galaxies; Sargsyan et al. 2014; but see D\'{i}az-Santos et al. 2017 for LIRGs) and \NII\ lines would have an uncertainty of a factor of 2 (e.g., Zhao et al. 2013, 2016). Meanwhile, one can employ the steep dependence of the \CII/\midco\ (or \NII/\midco) luminosity ratio on \fircolor\ to derive \fircolor\ (i.e., the dust temperature $T_{\rm dust}$; Lu15), which is essentially related to the average SFR surface density, $\Sigma_{\rm SFR}$ (Liu et al. 2015; Lutz et al. 2016), another fundamental parameter characterizing star formation in galaxies. Alternatively, one can estimate \fircolor\ from the \CI~370~\mum\ to \midco\ flux ratio (hereafter \CI/\midco; Lu et al. 2017a; Lu et al. 2018), which has the advantage that both lines can be observed simultaneously within one side frequency band (4 GHz) of ALMA since they are only separated by 2.7 GHz (rest-frame) in frequency. 

In addition, these lines are among the most luminous gas cooling lines, and thus important for cooling the interstellar medium (ISM). Further, they probe different gas phases in galaxies. The \CI\ and CO\,(1$-$0) lines come mainly from neutral medium (e.g., photon dominated regions; PDRs; Tielens \& Hollenbach 1985) with relatively low density since 
their critical densities ($n_{\rm crit}$) are only $\sim$10$^3$ cm$^{-3}$.
Recent observations suggest that the \CI\ lines can trace the total molecular gas as the CO\,(1$-$0) line does (Jiao et al. 2017). Unlike the \CI\ and CO\,(1$-$0) lines, however, the \midco\ line traces the warm (excitation temperature $T_{\rm ex}=150$ K) and dense ($n_{\rm crit}$$\sim$10$^5$ cm$^{-3}$) molecular gas that is in close proximity to the location of current or very recent SF activity. As shown in Lu et al. (2014, 2017a), the spectral line energy distribution (SLED) of LIRGs is generally peaking around the \midco\ line. Although the \CII\ line is considered a primary tracer of PDRs (Tielens \& Hollenbach 1985), it can also arise from ionized gas since it only takes $\sim$11.3 eV to turn C into C$^+$. On the other hand, the \NII\ line comes exclusively from ionized gas, and traces mainly diffuse, warm ISM due to its low critical density (44~cm$^{-3}$; Oberst et al. 2006). In summary, these lines form a valuable set of extinction-free probes into the properties of SF and gas in galaxies, especially high-$z$ objects.

In this paper, we present our ALMA imaging of the \midco\ line in \lessj\ (also known as LESS~J033229.4$-$275619), which lies at $z=4.755$ in the Extended {\it Chandra} Deep Field South. It hosts a Compton-thick AGN as revealed by the {\it Chandra} data (Xue et al. 2011), and is identified as a submillimetre galaxy (SMG) using ALMA high-resolution continuum observations (Hodge et al. 2013). Its properties, including line emission (e.g. CO(2$-$1), Coppin et al. 2010; \CII, De Breuck et al. 2011, 2014; Gullberg et al. 2018; and \NII, Nagao et al. 2012) and continuum emission (Coppin et al. 2009; Biggs et al. 2011; Wardlow et al. 2011;  Gilli et al. 2014), have been studied extensively throughout X-ray to mm/submm bands. Our new ALMA observations of the \midco\ and \CI\ lines and the corresponding continuum emission at $\sim$370 \mum\ (rest-frame) provide new insights into its properties of the (warm and dense) molecular gas and star formation. The paper is organized as follows: we describe our observations and data reduction in Section 2, present our results and discussion in Section 3, and summarize briefly in the last section. Throughout the paper, we adopt a Hubble constant of $H_0=71~$km~s$^{-1}$~Mpc$^{-1}$, $\Omega_{\rm M} =0.27$, and $\Omega_\Lambda=0.73$ (Spergel et al. 2007). At $z=4.755$, the luminosity distance ($D_{\rm L}$) is 44,866~Mpc and 1$\arcsec$ corresponds to 6.6 kpc.

\section{Observations and Data Reduction}
The \midco\ imaging observations of \lessj\ were conducted in two equal-duration runs on March 4 and 26, 2016, using the Band 4 receivers of ALMA in the time division mode (TDM) with a velocity resolution of $\sim$33.5 \kms. In each observation, the four basebands (i.e.,``Spectral Windows"; SPWs 0-3) were centered at the sky frequencies of 140.15, 142.15, 128.15 and 130.11 GHz, respectively, each with a bandwidth of about 2 GHz. During the observations, 42 and 38 12-meter antennas were used respectively, with baselines ranging from 15 to 460 meters. The total on-source integration is 9 minutes. The pointing, phase, bandpass and flux calibrations were based on J0334$-$4008 and J0348$-$2749.

The data were reduced with CASA 4.5.3 (McMullin et al. 2007). The calibrated data sets were combined and cleaned using the Briggs weightings with robust = 0.0, and have nearly identical synthesized beams for the line and continuum emission, with the full width at half maximum (FWHM) of $1\arcsec.22\times0\arcsec.95$ ($1\arcsec.19\times0\arcsec.90$ for the continuum) and a position angle of $\sim$87$^\circ$. The continuum was measured using data in SPWs 0-3 by excluding the line emission channels. For the \midco\ and \CI\ line emission, the cube was generated using the data in SPW-0, which encompasses both lines (separated by only 0.47 GHz at the redshift of 4.755) with an effective bandpass of about 4000~\kms.

In order to increase the signal-to-noise ratio (S/N), we binned spectral cubes into channels with the width of $\delta {\rm v}=100$ \kms. The noise for these channel maps in \midco\ is about 0.3~mJy~beam$^{-1}$. For the continuum, the 1$\sigma$ rms noise is about 40~$\mu$Jy~beam$^{-1}$. The zeroth moment map is integrated over the LSR radio velocity range of ${\rm v}=-217-382$~\kms\ (zero velocity corresponding to 806.652/5.755=140.16542 GHz), which excludes the \CI\ emission in the datacube of 100~\kms\ channel maps, and its 1$\sigma$ rms noise is 81 mJy~\kms~beam$^{-1}$. All noise measurements were performed on the maps before the primary beam correction, and unless otherwise stated, flux measurements are based on the images after the primary beam correction, whereas all of the figures are produced using the results before the primary beam correction.

\section{Results and Discussion}

%
%
%
%
%

\begin{figure}[t]
\centering
\includegraphics[width=0.475\textwidth,bb =30 133 558 990]{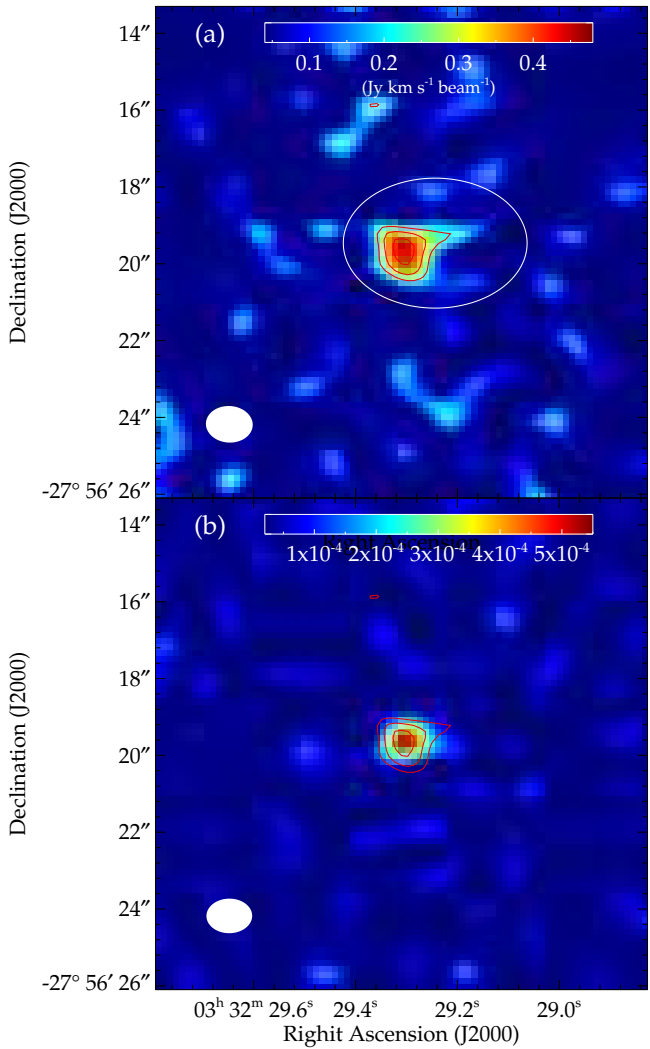}
\caption{\midco\ line emission contours of the integrated map superimposed on (a) the integrated \midco\ map, and (b) the 386\mum\ continuum. The contour levels are $[3, 4, 5] \times \sigma$ ($\sigma = 81$~Jy~\kms~beam$^{-1}$). The open ellipse in (a) gives the region used to measure the total flux and extract the spectrum. The filled ellipse in (a) and (b) show the beam shapes.}
\label{Figmom}
\end{figure}

\begin{figure*}[t]
\centering
\includegraphics[width=0.7\textwidth,bb =10 18 635 836]{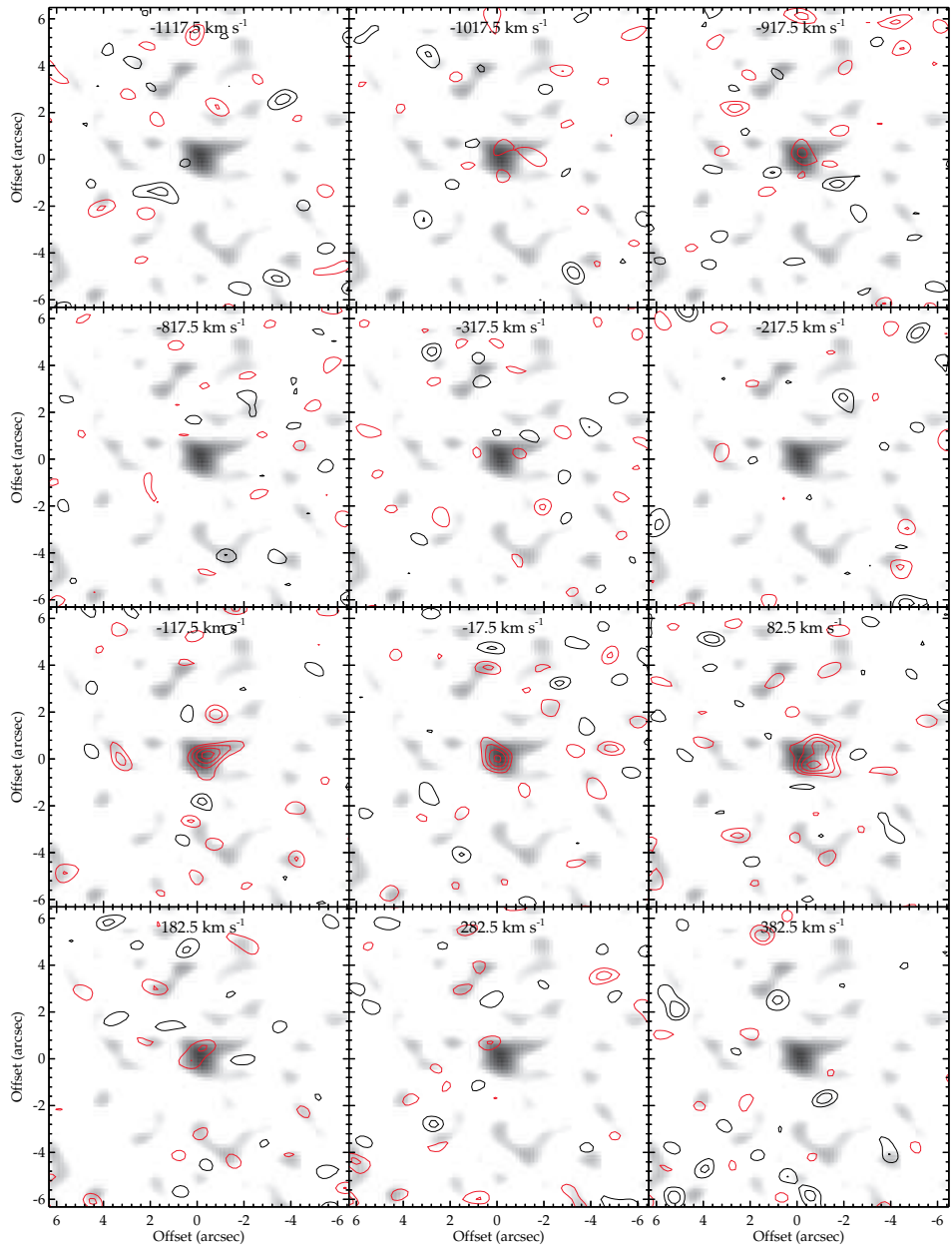}
\caption{\CI\ (the first four lowest velocity channels) and \midco\ line emission contours of the channel maps, overlaid on the integrated \midco\ emission image. The red contour levels are $[2, 3, 4, 5]\times\sigma$ ($\sigma$$\sim$0.3 mJy beam$^{-1}$). Equivalent negative contours are shown by black curves. In each channel, the central LSR (radio) velocity is labeled.}
\label{Figchan}
\end{figure*}

\begin{figure}[t]
\centering
\includegraphics[width=0.475\textwidth,bb = 100 22 646 616]{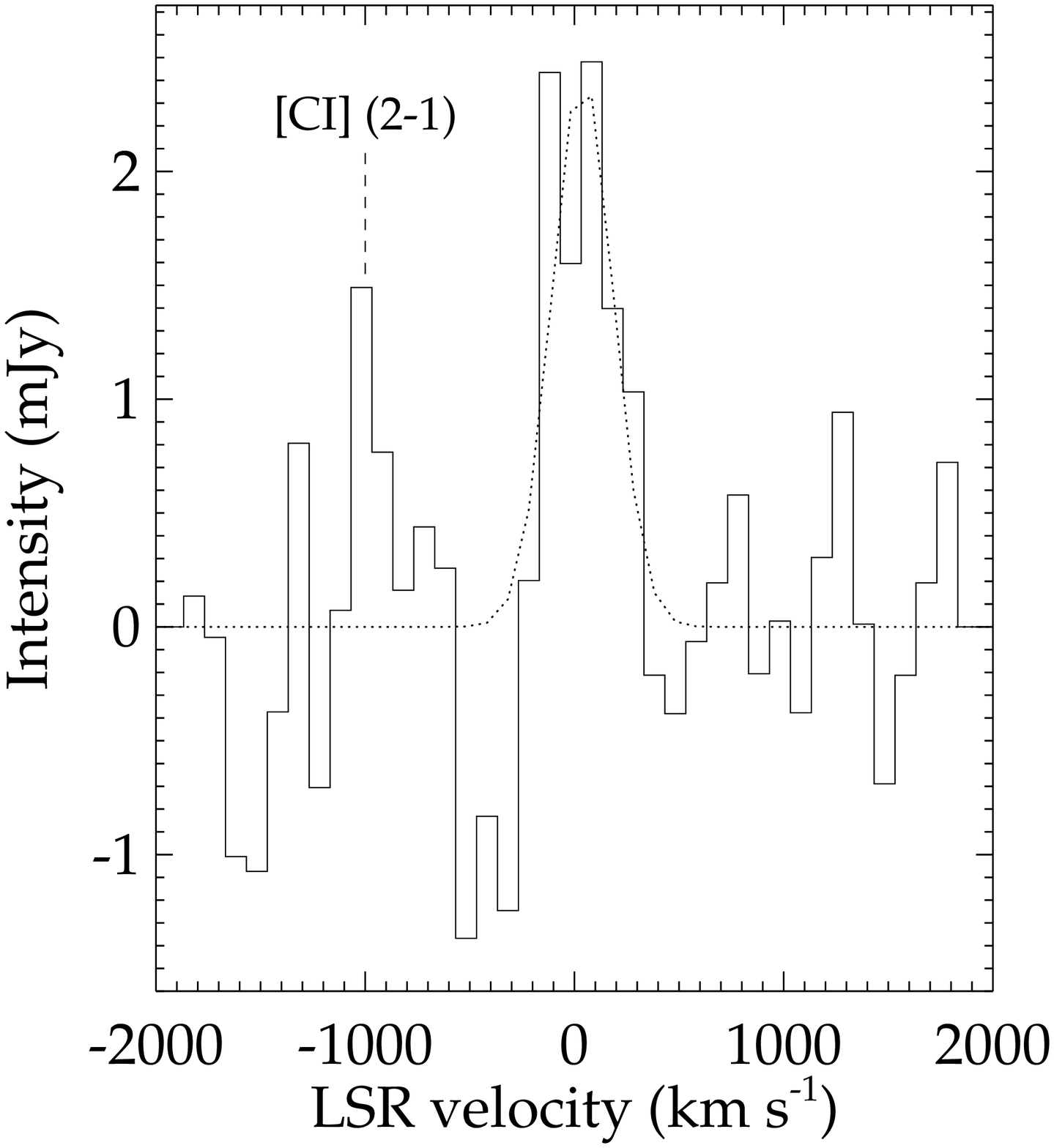}
\caption{Spatially integrated \midco\ line profile (after primary beam correction; solid line) extracted within the open ellipse in Figure 1. The velocity is relative to $z=4.755$. The dotted line gives the Gaussian fitting result, and vertical dashed line marks the expected position of the \CI\ line.}
\label{Figspec}
\end{figure}

\subsection{Line Emission}
\subsubsection{\midco\ Emission}
In Figure \ref{Figmom}a we show the integrated \midco\ emission map (i.e., the moment zero map) and its contours. The solid and open ellipses respectively give the beam and the region we used to extract the spectrum. The line emission is largely unresovled, with a deconvolved size (FWHM) of $1\arcsec.1(\pm0\arcsec.5)\times0\arcsec.9(\pm0\arcsec.8)$, which is very close to the beam size and thus suffers a large uncertainty (as given in the parentheses), and a highly uncertain position angle of $94\pm70^\circ$. 

From the intensity map we can see a tail towards the west. This tail only exists in the two channels of $v=-117.5$ and $82.5$~\kms, as shown in Figure \ref{Figchan}, which plots the channel maps. Unlike the \CII\ data presented in De Breuck et al. (2014), we can not see any no rotating pattern in the channel maps of the \midco\ emission. This might be due to the fact that both our spatial and velocity resolutions are much lower than that in De Breuck et al. (2014).

To obtain the integrated line flux ($f_{\rm CO\,(7-6)}$), we used three different methods: (1) Perform aperture photometry within the open ellipse shown in Figure \ref{Figmom}a, and we find $f_{\rm CO\,(7-6)}=0.83\pm0.04$~Jy~\kms; (2) Fit the 1-d spectrum (as shown in Figure \ref{Figspec}) with a Gaussian profile and we have $f_{\rm CO\,(7-6)}=0.89\pm0.09$~Jy~\kms; and (3) Use the CASA task {\it imfit} to fit the 2-d image and we obtain $f_{\rm CO\,(7-6)}=0.86\pm0.21$~Jy~\kms. All three fluxes are consistent well with each other, and thus we adopt the averaged $f_{\rm CO\,(7-6)}=0.86\pm0.08$~Jy~\kms\ as the final result. 

As shown in Figure \ref{Figspec}, our spatially integrated line profile has an FWHM of $343\pm40$~\kms, much wider than those from the CO (2$-$1) ($160\pm45$~\kms; Coppin et al. 2010) and \NII\ ($230\pm22$~\kms; Nagao et al. 2012) lines, but consistent with that from the \CII\ line ($\sim$350~\kms; De Breuck et al. 2014). The best-fit Gaussian profile also gives a central velocity of 39~\kms, indicating a redshift of 4.7558, in good agreement with those obtained from the CO~(2$-$1), \CII\ and \NII\ lines. 
 
\begin{figure}[t]
\centering
\includegraphics[width=0.475\textwidth,bb = 76 387 567 901]{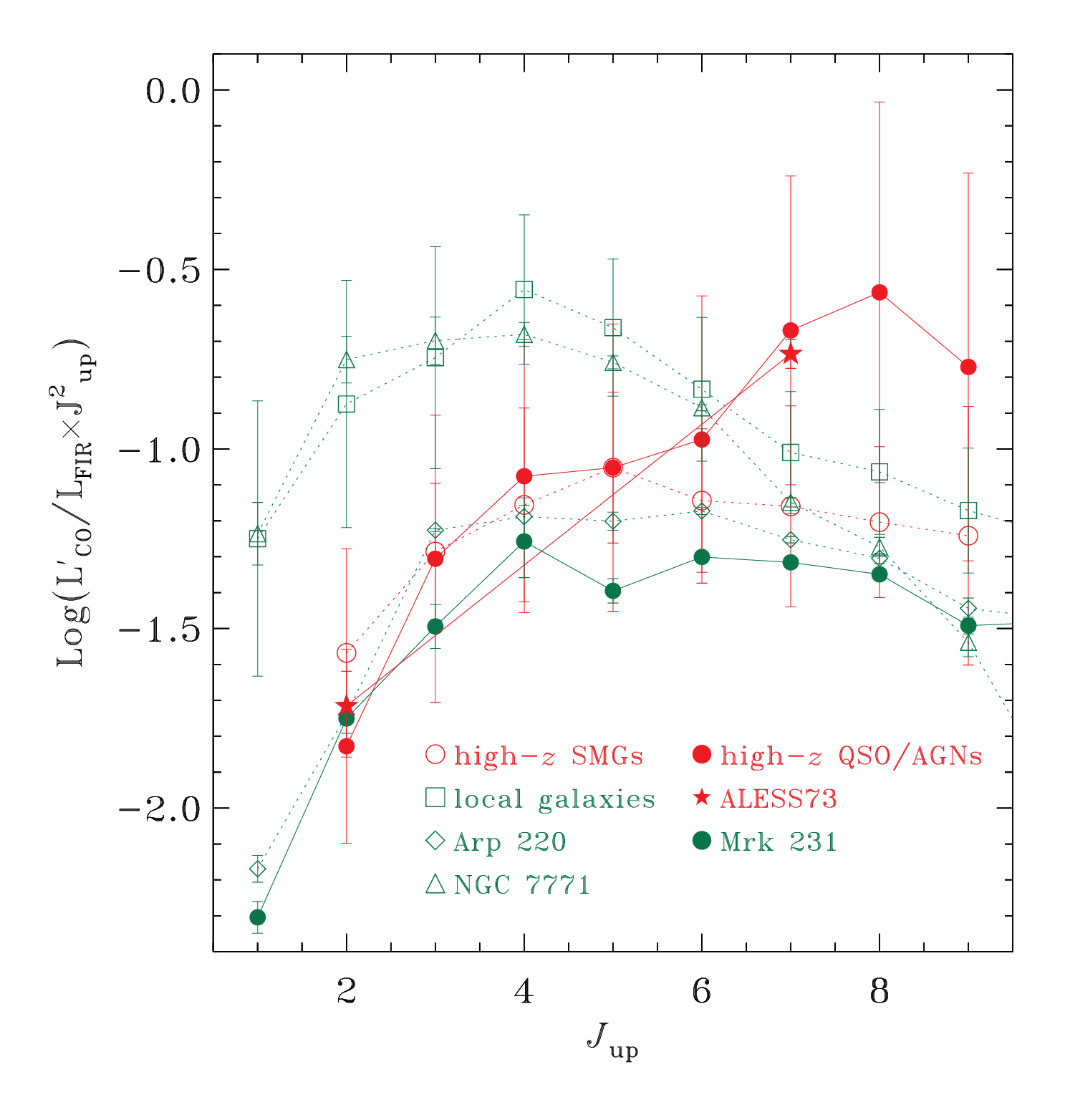}
\caption{FIR luminosity-normalised mean CO SLEDs for local (green symbols) and high-$z$ (red symbols) objects. $J_{\rm up}^2$ is included in {\it y}-axis so that the unit of {\it y}-axis is comparable to velocity integrated flux density. Local galaxies (squares) adopted from Liu et al. (2015) mostly have $10^9 L_\odot \leq L_{\rm IR} \leq 10^{12} L_\odot$). Here we also include three individual galaxies: a LIRG, NGC 7771 (Lu et al. 2017), a typical ULIRG, Arp220 (Rangwala et al. 2011; Lu et al. 2017), and an AGN-dominated ULIRG, Mrk 231 (van der Werf et al. 2010; Lu et al. 2017). High-redshift sources are from Carilli \& Walter (2013), Strandet et al. (2017), Yang et al. (2017), Venemans et al. (2017),  and Ca\~{n}ameras et al. (2018). We can see clearly that the CO SLEDs show different shapes: they peak at $J$$\sim$4 for local galaxies (including NGC 7771), remain flat ($4 \leq J \lesssim 9$) for the two ULIRGs Arp 220 and Mrk 231, and for high-$z$ SMGs, and show peak at $J$$\sim$7$-$9 for high-$z$ AGN/QSOs. }
\label{FigSLED}
\end{figure}

Following Solomon \& Vanden Bout (2005), we calculate the line luminosity and obtain $L^\prime _{\rm CO\,(7-6)}=(1.54\pm0.16)\times10^{10}$~K~\kms~pc$^2$. Comparing our $L^\prime _{\rm CO\,(7-6)}$ with $L^\prime _{\rm CO\,(2-1)}$ from Coppin et al. (2010), we find that the brightness temperature ratio of \midco/CO\,(2$-$1) of \lessj\ is $\sim$$0.8\pm0.2$, indicating that the CO emission at $J=7$ is nearly thermalized. To have a direct view of the CO excitation of different types of galaxies, we plot the mean CO SLEDs for local (Liu et al. 2015) and high-$z$ (Carilli \& Walter 2013; Strandet et al. 2017; Yang et al. 2017; Venemans et al. 2017; Ca\~{n}ameras et al. 2018) objects, as well as three individual galaxies, NGC~7771 (Lu et al. 2017), Arp~220 (Rangwala et al. 2011; Lu et al. 2017) and Mrk~231 (van der Werf et al. 2010; Lu et al. 2017), in Figure \ref{FigSLED}. Obviously, high-$z$ QSO/AGNs, including \lessj, have the highest excitation CO SLED peaking at $J$$\sim$$7-9$, which agrees well with the compact region and high SFR in their host galaxies, and/or the harder radiation field from the central AGNs. 

\subsubsection{\CI\ Emission}

In Figure \ref{Figspec} we also marked the expected position of the \CI\ emission. Indeed, there exists a peak with a S/N$\sim$4. In the channel maps we can also see these two channels (the top right two panels in Figure \ref{Figchan}). To estimate the total flux of \CI, we simply sum the two channels and get $f_{\rm [C\,{\scriptsize \textsc{i}}]}=0.23\pm0.08$~Jy~\kms, yielding a line luminosity of $L^\prime _{\rm [C\,{\scriptsize \textsc{i}}]}=(4.0\pm1.4)\times10^{9}$~K~\kms~pc$^2$. 

Adopting the CO-based \CI-to-total gas conversion factor of 12.5~$M_\odot$~(K~\kms~pc$^2$)$^{-1}$ in Jiao et al. (2017), we estimate the total molecular gas mass $M_{\rm H_2,CI}=(5.0\pm1.8)\times10^{10}$~$M_\odot$. Our derived $M_{\rm H_2,CI}$ is consistent within $3\sigma$ uncertainties with that ($M_{\rm H_2,CO}=(1.6\pm0.3)\times10^{10}$~$M_\odot$) derived in Coppin et al. (2010) using the CO~(2$-$1) luminosity. Following Scoville et al. (2016), we further calculate the total molecular gas mass using the rest-frame 850 \mum\ flux density from our fit to the FIR spectral energy distribution (SED; see the following section for detail), and obtain $M_{\rm H_2,cont}=(2.4\pm0.7)\times10^{11}$~$M_\odot$, which is $\sim$$5-15$ times higher than $M_{\rm H_2,CI}$ and $M_{\rm H_2,CO}$. We note that Scoville et al. (2016) adopted a CO-to-H$_2$ conversion factor ($\alpha_{\rm CO}$) of 6.5~$M_\odot$~(K~\kms~pc$^2$)$^{-1}$, while $\alpha_{\rm CO}=0.8$~$M_\odot$~(K~\kms~pc$^2$)$^{-1}$, which is usually used for ultra-luminous infrared galaxies and SMGs (e.g., Downes \& Solomon 1998), was adopted in the former two estimates. Therefore, taking into account the large difference in $\alpha_{\rm CO}$, these three independent estimates are consistent with each other. Here we adopt the average value of $M_{\rm H_2,CI}$ and $M_{\rm H_2,CO}$, i.e., $M_{\rm H_2}=(3.3\pm1.7)\times10^{10}$~$M_\odot$ as the final total molecular gas mass.

\subsection{Continuum Emission}
\begin{figure}[t]
\centering
\includegraphics[width=0.475\textwidth,bb = 65 366 562 696]{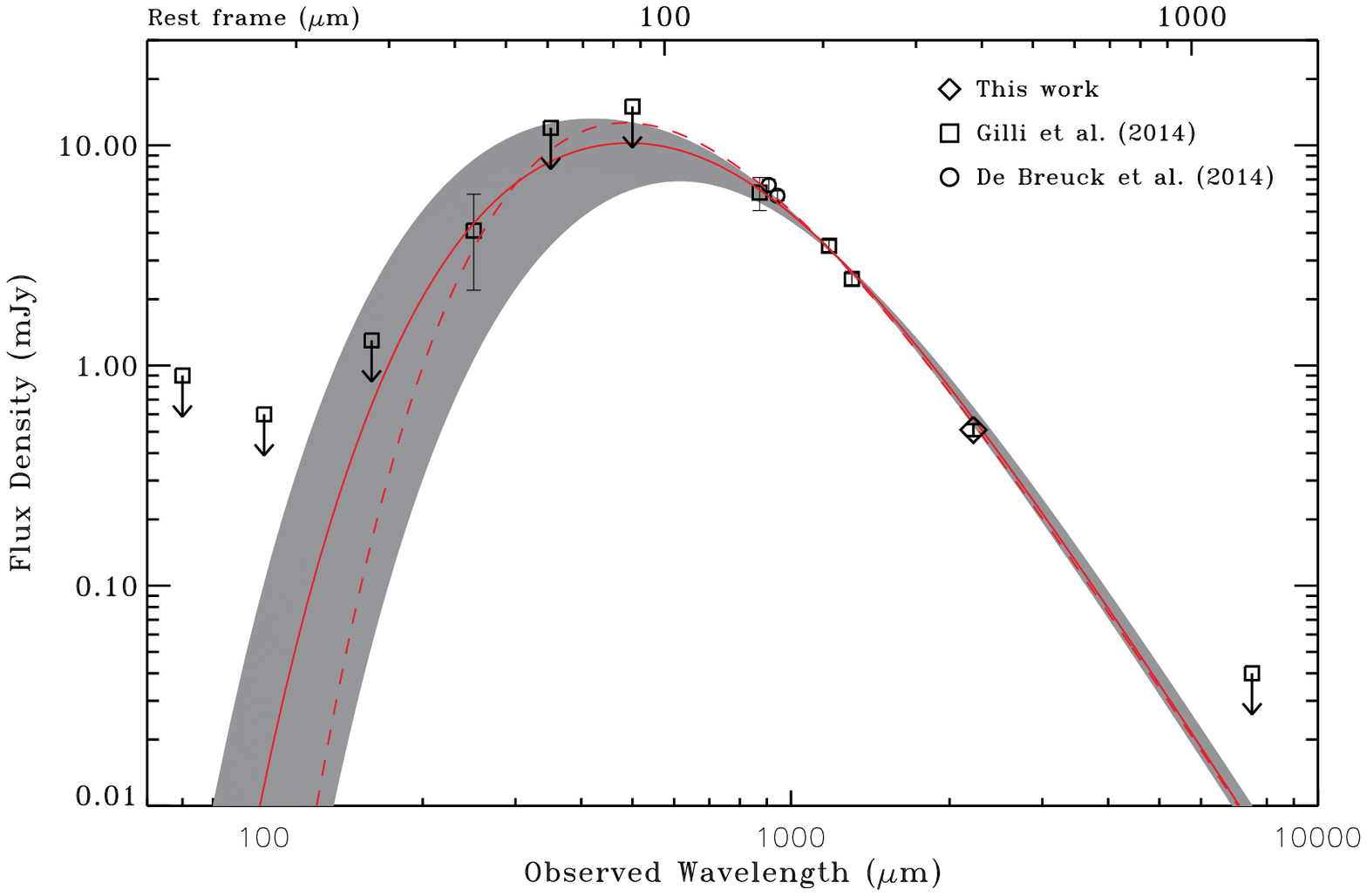}
\caption{The FIR SED (observed frame) of \lessj, with the corresponding rest-frame wavelength marked on the top axis. The best-fitted results for the optically thick and thin cases are shown by the red thick solid and dashed lines, respectively. The gray lines (i.e., the shading region) give all model results at the 95\% confidence level for the optically thick case.}
\label{Figsed}
\end{figure}

The continuum (rest-frame 385 \mum) map is shown in Figure \ref{Figmom}b with the \midco\ contours overlaid. Obviously our observation does not resolve the continuum emission. This seems not surprising: The continuum emission at 158 \mum, which is very marginally resolved (beam size: $0\arcsec.64\times0\arcsec.44$; De Breuck et al. 2014), only has a FWHM size of $0\arcsec.29 \pm 0.06$ ($1.9 \pm 0.4$~kpc), and further high-resolution ($0\arcsec.16)$ observations at rest-frame $\sim$150~\mum\ show that \lessj\ has a FWHM size of $0\arcsec.34 \pm 0\arcsec.02$ ($2.2 \pm 0.1$~kpc); though the continuums at $\sim$$150-160$ and 385 \mum\ could have different sizes. Measured within the same elliptical aperture as shown in Figure \ref{Figmom}a ($a=1\arcsec.9$ and $b=1\arcsec.3$), the flux density of the continuum is $0.51\pm0.02$~mJy.

Figure \ref{Figsed} plots the FIR spectral energy distribution (SED) of \lessj, where we have added a calibration uncertainty of 5\%, following Bonato et al. (2018), for ALMA photometric points that no calibration uncertainties were reported in the original references. Apart from the new data presented in this work, the remained photometric points are collected from the literature (De Breuck et al. 2014; Gilli et al. 2014 and see references therein). To fit the observed SED, we adopt a general single-temperature-modified blackbody, i.e.,
\begin{equation}
  S(\nu_{\rm obs}) \propto (1-e^{-\tau(\nu)})B(\nu, T_{\rm dust})
\end{equation}
where $B(\nu,T)$ is the Planck function, and $\tau(\nu) = (\nu/\nu_0)^\beta$ is the optical depth, $\beta$ is the dust emissivity power law index and $\nu_0$ is the rest frequency where the dust becomes optically thick (i.e. $\tau(\nu_0)= 1$). In the optically thin case, equation (1) simplifies to $S\nu~\propto~\nu^\beta B(\nu, T_{\rm dust})$.

To take the upper limits into account, we adopt the method proposed in Sawicki (2012) during the fitting process. We further fix $\beta=1.8$ (Planck Collaboration et al. 2011) due to our limited data points. For the general optically thick case, $\nu_0$ is set to be 1.5 THz (i.e. $\lambda_0=c/\nu_0=200$~\mum) according to the measurements in a number of local and distant starburst galaxies (e.g., Conley et al. 2011; Rangwala et al. 2011; Fu et al. 2012). For the optically thin model, data points with observed-frame wavelength of 160 \mum\ and shortward, which are clearly on the Wien tail where small grains likely dominate the emission, are excluded from the fitting process. As shown by the red lines in Figure \ref{Figsed}, both the general optically thick and the optically thin models can fit the observed SED well. In the optically thick case, however, the best-fit $T_{\rm dust}$ ($57.6\pm3.5$) is much higher than that ($35.8\pm1.9$) from the optically thin model.

The optically thick model yields the total infrared luminosity \LIR\ of $(5.8\pm0.9)\times10^{12}$~$L_\odot$, the FIR (42.5$-$122.5~\mum) luminosity \LFIR\ of $(4.0\pm0.6)\times10^{12}$~$L_\odot$, and \fircolor\ of $0.86\pm0.08$, whereas the optically thin model gives similar values (see Table \ref{Tabderived}). Using \LIR\ and $T_{\rm dust}$ in the Stefan$-$Boltzmann law, we obtain a spherical source radii of $R_{\rm SB,\,thick}=0.54\pm0.03$~kpc and $R_{\rm SB,\,thin}=1.4\pm0.2$~kpc, respectively for the optically thick and thin models. Bearing in mind that we have assumed that the filling factor of the blackbody (the starburst in this case) is equal to 1 when using the Stefan$-$Boltzmann law, however, this situation is almost never true since star-forming regions always break up in many clumps when looked at sufficient angular resolution. Further, the total infrared luminosity could be underestimated as we only fit the FIR part of the IR SED. Therefore, $R_{\rm SB,\,thick}$ and $R_{\rm SB,\,thin}$ should be lower limit of the true size.

As mentioned at the beginning of this section, the observed size (Gaussian FWHM) of the continuum emission at observed-frame 870 \mum\ is $2.2\pm0.1$~kpc (Hodge et al. 2016), which is in between $R_{\rm SB,\,thick}$ and $R_{\rm SB,\,thin}$. Given the fact that the sizes derived from the Stefan$-$Boltzmann law are lower limits, it seems that the source radius is more consistent with the estimation from the optically thick model, and thus we adopt the best-fit results from this model in the following analysis.

\begin{deluxetable}{lccccc}[t]
\centering
\tablecaption{Derived Physical Parameters from the SED Fitting}
\tablewidth{0pt}
\tabletypesize{\scriptsize}
\tablehead{
\colhead{\multirow{2}{*}{Model}}&\colhead{\multirow{2}{*}{$\chi^2_{\rm red}$}}&\colhead{$T_{\rm dust}$}&\colhead{\multirow{2}{*}{\fircolor}} & \colhead{$\log L_{\rm IR}$}&\colhead{$\log L_{\rm FIR}$}\\
&&\colhead{(K)}& & \colhead{($L_\odot$)}&\colhead{($L_\odot$)}
}
\startdata
Thick  &1.87  & $57.6(3.5)$ &$0.86(0.08)$&$12.76(0.07)$&$12.60(0.06)$\\
Thin   &1.42 & $35.8(1.9)$ &$0.81(0.12)$&$12.76(0.10)$&$12.65(0.09)$
\enddata
\tablecomments{Numbers in the parentheses give the $1\sigma$ uncertainties.}
\label{Tabderived}
\end{deluxetable}

\subsection{Line Ratios}
Table \ref{Tabratio} lists four line ratios, e.g., \CII-, \NII- and \CI-to-\midco\ ratios, and the \NII-to-\CII\ ratio, along with the derived \fircolor. We have used the equations in Lu15, Lu et al. (2017b, 2018) to estimate \fircolor\ from the observed \NII/\midco, \CII/\midco\ and \CI/\midco, respectively. Further, we exploit the correlation between \NII/\CII\ and \fircolor\ to derive \fircolor\ using the observed \NII/\CII\ ratio, i.e.

\begin{equation}
  f_{60}/f_{100}=(-0.37\pm0.09)-(0.96\pm0.08)\log {\rm [N\,{\scriptsize \textsc{ii}}]}/{\rm [C\,{\scriptsize \textsc{ii}}]}
\end{equation}
This correlation is the result of a least-square bisector fit to the local (U)LIRG sample presented in Lu15, and has a scatter of 0.16 dex in \fircolor.

As shown in Table \ref{Tabratio}, the FIR colors derived with various line ratios have an average value of $1.05\pm0.09$, and are in agreement (within $1-2\sigma$) with that obtained from the SED fit. Combining this result with those presented in Lu15 and Lu et al. (2018), we can conclude that our line-ratio based method, which has been calibrated using the data of local (U)LIRGs, is able to efficiently estimate \fircolor\ in high-$z$ sources with good accuracy. 

We also note that the \midco-involved line ratios tend to give larger \fircolor. This is due to the fact that the observed line-to-\midco\ ratios are $0.3-0.4$ dex smaller than the average value of local (U)LIRGs at $f_{60}/f_{100}$$\sim$0.9, indicating a possible enhancement of the \midco\ emission in \lessj. We will further discuss this in the following section.

\begin{deluxetable}{lccc}[t]
\centering
\tablecaption{Line Luminosity Ratios}
\tablewidth{0pt}
\tabletypesize{\scriptsize}
\tablehead{
\colhead{Ratio}&\colhead{Value\tablenotemark{a}}&\colhead{\fircolor\tablenotemark{b}} & \colhead{Reference\tablenotemark{c}}
}
\startdata
\CII/\midco  &$1.30\pm0.05$  & $1.02\pm0.22$ &1\\
\NII/\midco  &$0.12\pm0.07$  & $1.10\pm0.15$ &2\\
\NII/\CII   &$-1.18\pm0.09$  & $0.76\pm0.16$ &2, 1\\
\CI/\midco   &$-0.57\pm0.15$ & $1.31\pm0.18$ &3
\enddata
\tablenotetext{a}{Ratios are in logarithmic sacle.}
\tablenotetext{b}{FIR colors are derived with the relations presented in Lu15 and Lu et al. (2018).}
\tablenotetext{c}{Reference for line (excluding \midco) fluxes: (1) De Breuck et al. (2014); (2) Nagao et al. (2012); (3) this work.}
\label{Tabratio}
\end{deluxetable}

\subsection{Star Formation Properties}
\subsubsection{Estimation of the SFR}
To estimate the SFR of \lessj, we assume a Salpeter (1959) initial mass function and employ the three calibrators, i.e. \lco\, \LNII\ and \LFIR. The total infrared luminosity is not adopted here since we only fit the FIR part of the IR SED, and thus could underestimate \LIR.  Using equation (1) in Lu15, we have ${\rm SFR_{CO\,(7-6)}}=3300^{+1200}_{-900}~M_\odot~{\rm yr}^{-1}$. Since $f_{60}/f_{100}=0.86$, we adopt the \NII\ calibrator suitable for ``warm" (i.e., $0.6\leq f_{60}/f_{100}<0.9$) galaxies in Zhao et al. (2016), and obtain ${\rm SFR_{[N\,{\scriptsize\textsc{ii}}]}}=760^{+590}_{-330}~M_\odot~{\rm yr}^{-1}$ based on the \NII\ flux given in Nagao et al. (2012). Assuming $L_{\rm FIR}=L_{\rm IR}/1.9$ (e.g., Dale \& Helou 2002) and using the SFR calibrator given in Kennicutt (1998), we have ${\rm SFR_{FIR}}\sim1300\pm200~M_\odot~{\rm yr}^{-1}$, where the uncertainty from the calibrator itself is not taken into account.

\begin{figure}[t]
\centering
\includegraphics[width=0.475\textwidth,bb = 60 372 330 632]{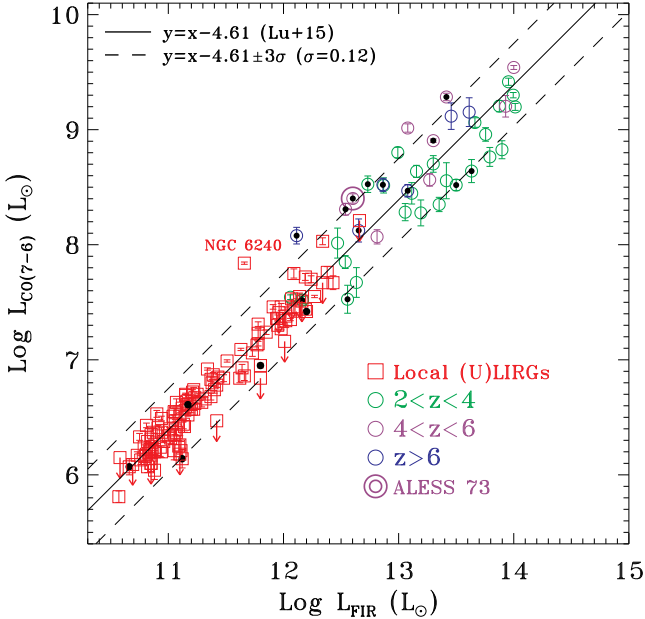}
\caption{FIR luminosity plotted against the \midco\ luminosity. Local (U)LIRGs (red squares) are adopted from Lu et al. (2015), and high-$z$ sources (circles) are compiled from  the literature (Solomon et al. 2005, Carilli \& Walter 2013, Strandet et al. 2017, Yang et al. 2017, Venemans et al. 2017,  and Ca\~{n}ameras et al. 2018). Sources with an additional solid dot are AGN/QSOs. The labeled galaxy NGC 6240 is a typical merger-driven shock-excited object, and has a much higher CO-to-FIR luminosity ratio. We also note that some sources are uncorrected for gravitational magnification.}
\label{Figco2fir}
\end{figure}

We can see that ${\rm SFR_{[N\,{\scriptsize\textsc{ii}}]}}$ and ${\rm SFR_{FIR}}$ is consistent with each other, whereas ${\rm SFR_{CO\,(7-6)}}$ is about $3-4$ times higher than the previous two SFRs, though they are still consistent with each other within $3\sigma$ uncertainties. The apparently larger SFR from the \midco\ line might also be attributed to the enhanced CO emission in \lessj. In fact, the \midco-to-FIR luminosity ratio \lco/\LFIR\ in \lessj\ is $\sim$$10^{-4.20\pm0.08}$, which is $>$$3\sigma$ higher than the mean value (10$^{-4.61\pm0.12}$) for local (U)LIRGs (Lu15). To further check whether \lessj\ is a true outlier or not, we plot \LFIR\ against \lco, both for local (U)LIRGs and for high-$z$ objects, in Figure~\ref{Figco2fir}. The solid and dashed lines are respectively the mean (${L_{\rm CO\,(7-6)}/L_{\rm FIR}}=10^{-4.61}$) and 3$\sigma$ ($\sigma=0.12$) values for local (U)LIRGs from Lu et al. (2015). Clearly, high-$z$ objects follow the local \LFIR-\lco\ relation (with mean ${L_{\rm CO\,(7-6)}/L_{\rm FIR}}=10^{-4.60}$), but have a significantly larger dispersion (0.29 dex). This larger dispersion in high-$z$ objects might mainly be due to the fact that they likely have a much larger error in their \LFIR\ as for many of them the FIR luminosity was either derived from a FIR-radio correlation or scaled from a sub-millimeter flux (see Carilli \& Walter 2013).

In spite of the small size of the samples, from Figure \ref{Figco2fir} we note that, unlike local strong AGNs, high-$z$ AGNs/QSOs tend to have a higher-than-average ${L_{\rm CO\,(7-6)}/L_{\rm FIR}}$ ratio ($10^{-4.49\pm0.34}$). Out of the 7 high-$z$ objects having $>3\sigma$ deviation from the local mean value, five are AGN/QSOs. This result seems not unreasonable since these high-$z$ objects might have mechanical heating from AGN-driven shocks and/or radio jets, which can enhance the gas emission (see Lu et al. 2014 for the local AGN-driven outflow galaxy NGC~1266, and for the radio jet galaxy 3C~293). Further, high-$z$ gas-rich quasars usually are believed to reside in merging system, which could also lead to SF-unrelated mechanical heating from galaxy-galaxy collision (e.g. NGC~6240 for a local example; Meijerink et a. 2013; Lu et al. 2014). 

Lesaffre et al. (2013) demonstrate that the CO emission can be more boosted than \CII, \NII\ and \CI, when low-velocity ($\sim$10 km~s$^{-1}$) shocks pass through a low-density ($10^2-10^4$ cm${^{-3}}$), mildly UV-irradiated (sufficient to ionize the carbon) molecular medium. This picture is consistent with our results. Furthermore, we calculated \fircolor\ using these line ratios for NGC~6240 and NGC~1266. We found that, similar to \lessj, the \midco-involved line ratios also predict larger \fircolor, whereas the \NII/\CII\ ratio gives \fircolor\ closer to the observed value. Therefore, it seems that SF-unrelated shocks tend to play a role in heating the gas in \lessj. In consequence, the combination of the \CII\ and \NII\ lines is likely a more robust method for estimating SFR and \fircolor\ for such system, though a much larger sample is needed to reach a solid conclusion. 

\subsubsection{The SFR Surface Density}
As shown in Liu et al. (2015) and Lutz et al. (2016), $\Sigma_{\rm SFR}$ is generally correlated both with \fircolor\ and $f_{70}/f_{160}$, with a scatter as large as $\sim$0.6 dex in $\Sigma_{\rm SFR}$. Using our best-fit result (e.g., $f_{60}/f_{100}=0.86$ and $f_{70}/f_{160}=1.71$), nevertheless, these two independent correlations give comparable estimates for $\Sigma_{\rm SFR}$ of $\sim$100~$M_\odot~{\rm yr}^{-1}~{\rm kpc}^{-2}$ for \lessj, with an uncertainty of a factor of 4. Further, we obtain $\Sigma_{\rm SFR}=100^{+78}_{-44}~M_\odot~{\rm yr}^{-1}~{\rm kpc}^{-2}$ using our ${\rm SFR_{[N\,{\scriptsize\textsc{ii}}]}}$ and the FIR size ($R_{150\mu {\rm m}}=1.1$~kpc) at rest-frame $\sim$150~\mum\ from Hodge et al. (2016), which is in good agreement with those derived from the $\Sigma_{\rm SFR}-$FIR color relations.

We can also have a rough estimation of $\Sigma_{\rm SFR}$ using the source size ($R_{\rm {SB,thick}}$$\sim$540~pc) derived in \S3.2, namely $\Sigma_{\rm SFR,SB}=(\frac{1}{2}{\rm SFR_{[{\rm N}\,\scriptsize \textsc{ii}]}}/(\pi R^2)\sim400~M_\odot~{\rm kpc}^{-2}$. These two $\Sigma_{\rm SFR}$ have a discrepancy of a factor of 4. However, the size from Stefan$-$Boltzmann law is a lower limit and the true $\Sigma_{\rm SFR}$ should be larger than $\Sigma_{\rm SFR,SB}$, which is indeed the case. The gas depletion time, $\tau_{\rm depletion}=M_{\rm gas}/{\rm SFR}=3.2\times10^{10}\,M_\odot/760\,M_\odot\,{\rm yr}^{-1} = 4.2\times10^7~{\rm yr}$ for \lessj, is similar to the gas depletion timescale for other SMGs (Carili \& Walter 2013).

\section{Summary}
In this paper we present our ALMA observations of the \midco\ line emission and the continuum emission at $\sim$386\,\mum\ in the young starburst galaxy \lessj\ at $z=4.755$. Our main results are:

\begin{enumerate}
	\item At the resolution of $1\arcsec.22\times0\arcsec.95$ ($1\arcsec.19\times0\arcsec.90$ for the continuum), the \midco\ emission is largely unresolved, and the continuum emission is totally unresolved. The deconvolved size of the \midco\ emission is $1\arcsec.1(\pm0\arcsec.5)\times0\arcsec.9(\pm0\arcsec.8)$. 
	\item The \midco\ line has a width of $343\pm40$~\kms. Its integrated flux is $0.86\pm0.08$~Jy~\kms, corresponding to a luminosity of $(2.5\pm0.3)\times10^{8}~L_\odot$. The \CI\ line has an integrated flux of $0.23\pm0.08$~Jy~\kms, corresponding to a luminosity of $(6.8\pm2.4)\times10^7~L_\odot$. For the continuum emission, it has a flux density of 0.51~mJy.
	\item By fitting the observed FIR SED of \lessj\ with a single-temperature modified blackbody, we obtain $T_{\rm dust}=57.6\pm3.5$~K, $f_{60}/f_{100}=0.86\pm0.08$, $L_{\rm FIR}=(4.0\pm0.6)\times10^{12}~L_\odot$, and $L_{\rm IR}=(5.8\pm0.9)\times10^{12}~L_\odot$, making it a luminous ULIRG.
	\item The SED-fit-based \fircolor\ is consistent with those derived from various line ratios. Therefore, our line-ratio based method, which is originally proposed in Lu15, can be practically used to derive \fircolor\ efficiently for high-$z$ sources. 
	\item The total molecular gas mass of \lessj\ is $(3.3\pm1.7)\times10^{10}~M_\odot$. Combing with the SFR estimated using \LNII, we obtain a depletion time of about 43~Myr.

\end{enumerate}

\begin{acknowledgements}
We thank the anonymous referee for her/his helpful comments. We also thank  Dr. Daizhong Liu and Dr. Chentao Yang for providing us their CO SLED data. The work is supported in part by the National Key R\&D Program of China grant No. 2017YFA0402704, the NSFC grant NOs. 11673057, 11673028, 11861131007, 11420101002), and Chinese Academy of Sciences Key Research Program of Frontier Sciences (grant No. QYZDJSSW-SLH008). This paper makes use of the following ALMA data: ADS/JAO.ALMA\#2015.1.00388.S. ALMA is a partnership of ESO (representing its member states), NSF (USA), and NINS (Japan), together with NRC (Canada) and NSC and ASIAA (Taiwan), in cooperation with the Republic of Chile. The Joint ALMA Observatory is operated by ESO, AUI/NRAO, and NAOJ. This research has made use of the NASA/IPAC Extragalactic Database (NED), which is operated by the Jet Propulsion Laboratory, California Institute of Technology, under contract with the National Aeronautics and Space Administration.
\end{acknowledgements}


\end{document}